\documentstyle[12pt,twoside,fleqn,espcrc1,epsfig]{article}

\def\beq{\begin{equation}}
\def\eeq{\end{equation}}
\def\bea{\begin{eqnarray}}
\def\eea{\end{eqnarray}}
\def\bce{\begin{center}}
\def\ece{\end{center}} 
\def\eg{{\it e.g.}}
\def\ie{{\it i.e.}}
\def\etal{{\it et al.}}

\def\rhs{{\it r.h.s.}}

\title{Vector Mesons in Medium and Dileptons in Heavy-Ion Collisions}

\author{Ralf Rapp\address{Department of Physics and Astronomy,
        SUNY Stony Brook, New York 11794-3800, USA} 
        \thanks{research supported in part by the A.-v.-Humboldt Foundation 
        (Feodor-Lynen program) and U.S. Department of Energy under Grant No. 
         DE-FG02-88ER40388.}  
        and Jochen Wambach\address{Institut f\"ur Kernphysik, TU Darmstadt,
        Schlo{\ss}gartenstr. 9, D-64289 Darmstadt, Germany}
        }

\begin{document}
\maketitle

\begin{abstract}
Theoretical approaches to assess modifications of vector mesons in the 
medium, as well as their experimental identification via electromagnetic 
probes, are discussed. Implications for the nature of chiral symmetry 
restoration in hot/dense matter are outlined and put into context with 
the axialvector channel.   
\end{abstract}

\section{Introduction}

The investigation of hadron properties inside atomic nuclei constitutes 
one of the traditional research objectives in nuclear physics. However,  
in terms of the underlying theory of strong interactions  
QCD) even the description of the nuclear ground state 
remains elusive so far. 
Valuable insights can be expected from a careful 
study of transition regimes between hadronic and quark-gluon degrees 
of freedom. {\it E.g.}, in electron-nucleus scattering experiments the
corresponding control variable is the momentum transfer, whereas heavy-ion 
reactions, performed over a wide range of collision energies, aim at   
compressing and/or heating normal nuclear matter to witness potential
phase transitions into a Quark-Gluon Plasma (QGP).  

Among the key properties of the low-energy sector of strong interactions 
is the (approximate) chiral symmetry of the QCD Lagrangian and its spontaneous
breaking in the vacuum. This is evident from such important
phenomena as the build-up of a chiral condensate and constituent
quark mass ($M_q\simeq 0.4$~GeV), or the large mass splitting
of $\Delta M \simeq 0.5$~GeV between 'chiral partners' in the hadron spectrum 
(such as $\pi(140)$-$\sigma(400-1200)$, $\rho(770)$-$a_1(1260)$ or 
$N(940)$-$N^*(1535)$).
It also indicates that medium modi\-fications of hadron properties can  
be viewed as precursors of chiral symmetry restoration. 

In this talk the focus will be on the vector (V) and axialvector (A) channels.  
The former is special in that it directly couples to the electromagnetic
current (\ie, real and virtual photons)
at which point it becomes 'immune' to (strong) final
state interactions thus providing direct experimental access to in-medium 
properties of vector mesons, {\it e.g.}, through photoabsorption/-production
on nuclei, or dilepton ($e^+e^-$, $\mu^+\mu^-$) spectra in heavy-ion 
reactions. The key issue is then
to relate the medium effects to mechanisms of chiral restoration. This 
necessitates the simultaneous consideration of the axialvector channel,  
which, however, largely has to rely on theoretical analyses.   

This talk is structured as follows: 
Sect.~2 is devoted to vector-meson properties in nuc\-lear matter, 
Sect.~3 contains applications to heavy-ion reactions and Sect.~4 finishes 
with conclusions.
A more complete discussion of the presented topics can be found in a
recent review~\cite{RW00}.

\section{(Axial-) Vector Mesons in Cold Nuclear Matter}

\subsection{Correlators and Duality Threshold}

The general quantity that is common to most theoretical approaches is
the current-current correlation function which in the (axial-) vector
channel is defined by 
\beq
\Pi^{\mu\nu}_{V,A}(q)=-i\int d^4x \ e^{iqx} \  
\langle 0| {\cal T} j_{V,A}^\mu(x) j_{V,A}^\nu(0)|0\rangle \ .  
\eeq
For simplicity we will concentrate on the (prevailing) isospin $I=1$ 
(isovector) projections 
\beq
j^\mu_{I=1}=\frac{1}{2}(\bar u~\Gamma^\mu~u-\bar d~ \Gamma^\mu~ d)
\quad  {\rm with} \quad 
\Gamma_V^\mu=\gamma^\mu \ , \ \ \Gamma_A^\mu=\gamma_5\gamma^\mu \ . 
\eeq
At sufficiently high invariant mass 
both correlators can be described by their (identical) perturbative
forms which read (up to $\alpha_S$ corrections)
\beq
{\rm Im}~\Pi_{V,I=1}^{\mu\nu} ={\rm Im}~\Pi_{A,I=1}^{\mu\nu}=  
-(g^{\mu\nu}-\frac{q^\mu q^\nu}{M^2}) \frac{M^2}{12} \ \frac{N_c}{2} \ , 
 \  \ M\ge M_{dual} \  
\eeq
($M^2=q_0^2-\vec q^2$).
At low invariant masses the vector correlator is 
accurately saturated by the (hadronic) $\rho$ spectral function within the
Vector Dominance Model (VDM), \ie, 
\bea
{\rm Im}~\Pi_{V,I=1}^{\mu\nu} &=& \frac{(m_\rho^{(0)})^4}{g_\rho^2} \ 
{\rm Im}~D_\rho^{\mu\nu} \ ,  \qquad\qquad\qquad\qquad\qquad\qquad   
M\le M_{dual} 
\\
{\rm Im}~\Pi_{A,I=1}^{\mu\nu} &=& \frac{(m_{a_1}^{(0)})^4}{g_{a_1}^2} \ 
{\rm Im}~D_{a_1}^{\mu\nu} - f_\pi^2 \ \pi \  \delta(M^2-m_\pi^2) \ 
q^\mu q^\nu  \ , \  \    M \le M_{dual}  
\eea 
with a similar relation involving the $a_1$ meson in the axialvector channel. 
The spontaneous breaking of chiral symmetry (SBCS) manifests itself in both 
the difference of the $a_1$ and $\rho$ spectral functions as well as 
the additional pionic piece
in $\Pi_A$ (notice that $f_\pi$ is another order parameter of SBCS).
In vacuum the transition from the hadronic to the partonic regime 
('duality threshold') is characterized by the onset 
of perturbative QCD around $M_{dual}\simeq 1.5$~GeV. 
In the medium, chiral restoration requires the degeneration of 
$V$- and $A$-correlators over the entire mass range.

\subsection{Model-Independent Results: 
V-A Mixing and Sum Rules}
In a dilute gas the prevailing  medium effect can be computed via 
low-density expansions. Using soft pion theorems and current algebra
Krippa~\cite{Krip98} extended an earlier 
finite-temperature analysis~\cite{DEI90} to the 
finite-density case to obtain
\bea
\Pi^{\mu\nu}_V(q) &=& (1-\xi) \ \Pi^{\circ\mu\nu}_V(q)
+\xi \ \Pi^{\circ\mu\nu}_A(q)
\nonumber\\
\Pi^{\mu\nu}_A(q) &=& (1-\xi) \ \Pi^{\circ\mu\nu}_A(q)
+\xi \ \Pi^{\circ\mu\nu}_V(q) \ ' 
\label{vamix}
\eea  
\ie, the leading density effect is a mere 'mixing' of the vacuum 
correlators $\Pi^{\circ\mu\nu}$. The 'mixing' parameter
\beq
\xi\equiv\frac{4\varrho_N \bar{\sigma}_{\pi N}} {3f_\pi^2 m_\pi^2}  
\label{xi}
\eeq
($\varrho_N$: nucleon density) is determined by the 'long-range' part 
of the $\pi N$ sigma term, 
\beq
\bar{\sigma}_{\pi N} = 4\pi^3 m_\pi^2 \langle N|\pi^2|N \rangle \
\simeq 20~{\rm MeV} \ . 
\eeq
Chanfray \etal~\cite{CDER98} pointed out that $\bar{\sigma}_{\pi N}$ 
is in fact governed by the  
well-known nucleon- and delta-hole excitations in the pion cloud 
of the $\rho$ (or $a_1$) meson which have been thoroughly studied
within hadronic models to be discussed in the following section. 
A naive extrapolation of eq.~(\ref{xi}) to the chiral restoration point 
where $\xi =1/2$, gives $\varrho_c\simeq 2.5\varrho_0$, which is not 
unreasonable. Nonetheless, 
as we will see below, realistic models exhibit substantial medium 
modifications beyond the mixing effect. 

Similar in spirit, \ie, combining low-density expansions with chiral
constraints, is the so-called master formula approach  
applied in ref.~\cite{SYZ}: chiral Ward identities including 
the effects
of explicit breaking are used to express medium corrections to
the correlators through empirically inferred $\pi N$, $\rho N$ 
(or $\gamma N$, etc.) scattering amplitudes times density.  Resummations
to all orders in density cannot be performed either in this framework. 
 
Model independent relations which are in principle
valid to all orders in density are provided by sum rules.  
Although typically of little predictive power, their 
evaluation in model calculations can give valuable insights.   
One example are the well-known QCD sum rules which have been 
used to analyze vector-meson spectral functions in 
refs.~\cite{KKW97,LPM98}. It has been found, \eg, that the generic 
decrease of the quark- and gluon-condensates on the right-hand-side 
(\rhs) is compatible with the phenomenological (left-hand) side if 
{\em either} (i) the vector meson masses decrease (together with small
resonance widths),   {\em or}, (b) 
both  width and mass increase (as found in most microscopic models). 

Another example of sum rules are the ones derived by 
Weinberg~\cite{Wei67}, being generalized to the in-medium 
case in ref.~\cite{KS94}.
The first Weinberg sum rule, \eg, connects the pion decay constant to 
the integrated difference between the $V$- and $A$-correlators:
\beq
f_\pi^2=-\int\limits_0^\infty \frac{dq_0^2}{\pi (q_0^2-q^2)} 
\left[ {\rm Im}~\Pi_V(q_0,q)-{\rm Im}~\Pi_A(q_0,q)\right] 
\label{wsr1}
\eeq
for arbitrary three-momentum $q$ 
(here, the pionic piece has been explicitly separated out from  
$\Pi_A$). We will come back to this relation below.

\subsection{Hadronic Models and Experimental Constraints}

Among the most spectacular predictions for the behavior of vector mesons 
in medium is the Brown-Rho Scaling hypothesis~\cite{BR91}. By imposing 
QCD scale invariance on a chiral effective Lagrangian at finite density
and applying a  mean-field approximation it was conjectured that all
light hadron masses (with the exception of the symmetry-protected 
Goldstone bosons) drop with increasing density following an approximately
universal scaling law. The scaling also encompasses the pion decay constant
(as well as an appropriate power of the quark condensate) and therefore
establishes a direct link to chiral symmetry restoration being realized
through the vanishing of all light hadron masses.

\begin{figure}[!htb]
\begin{minipage}[t]{78mm}
\hspace{-0.4cm}
\epsfig{file=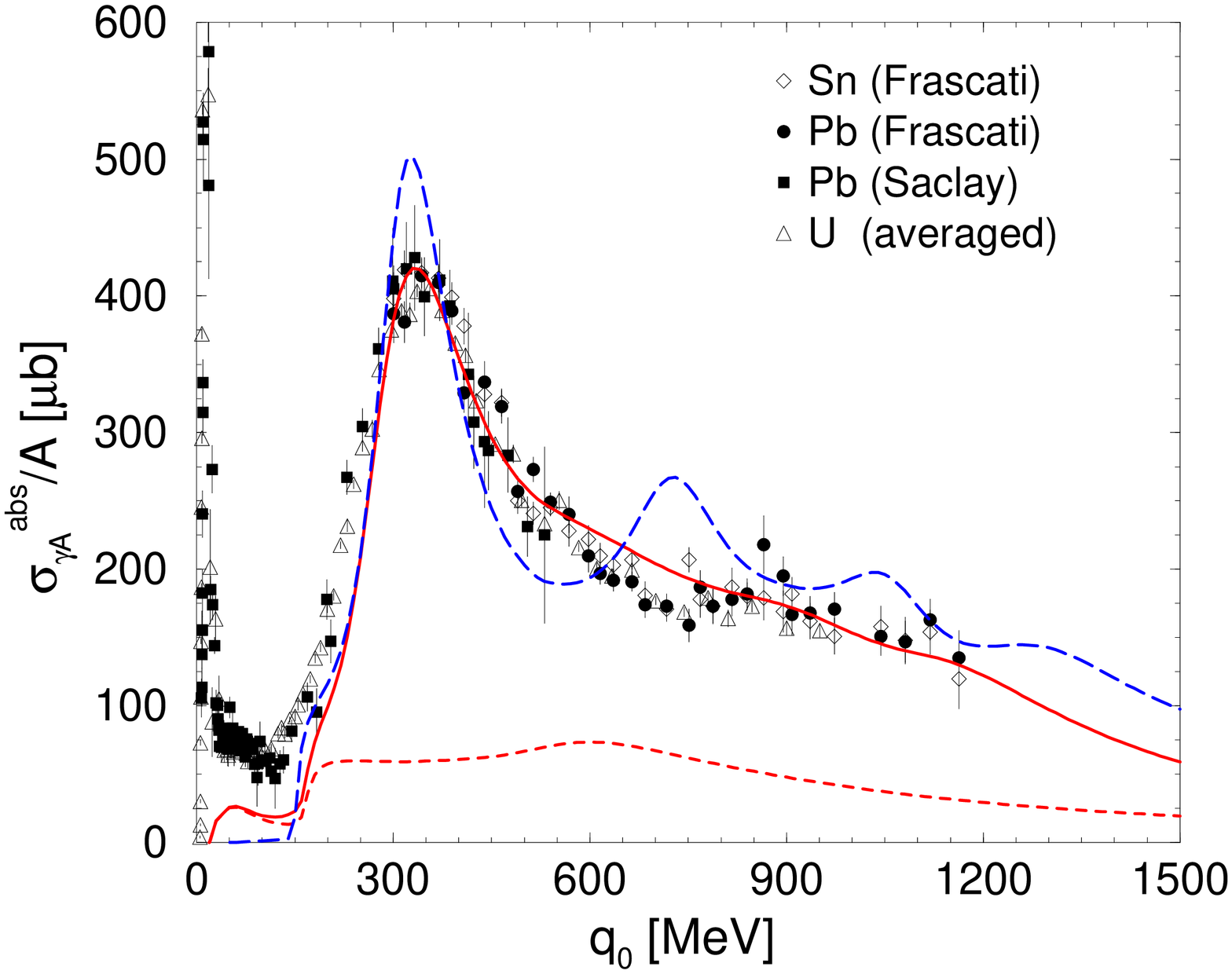,width=6.6cm,angle=-90}
\vspace{-1cm}
\caption{Photoabsorption spectra on nuclei with (full line) and
without (long-dashed line) higher order medium effects~\cite{RUBW};
short-dashed line: $\Sigma_{\rho\pi\pi}$ contribution.}
\label{fig_phabs}
\end{minipage}
\hspace{\fill}
\begin{minipage}[t]{78mm}
\epsfig{file=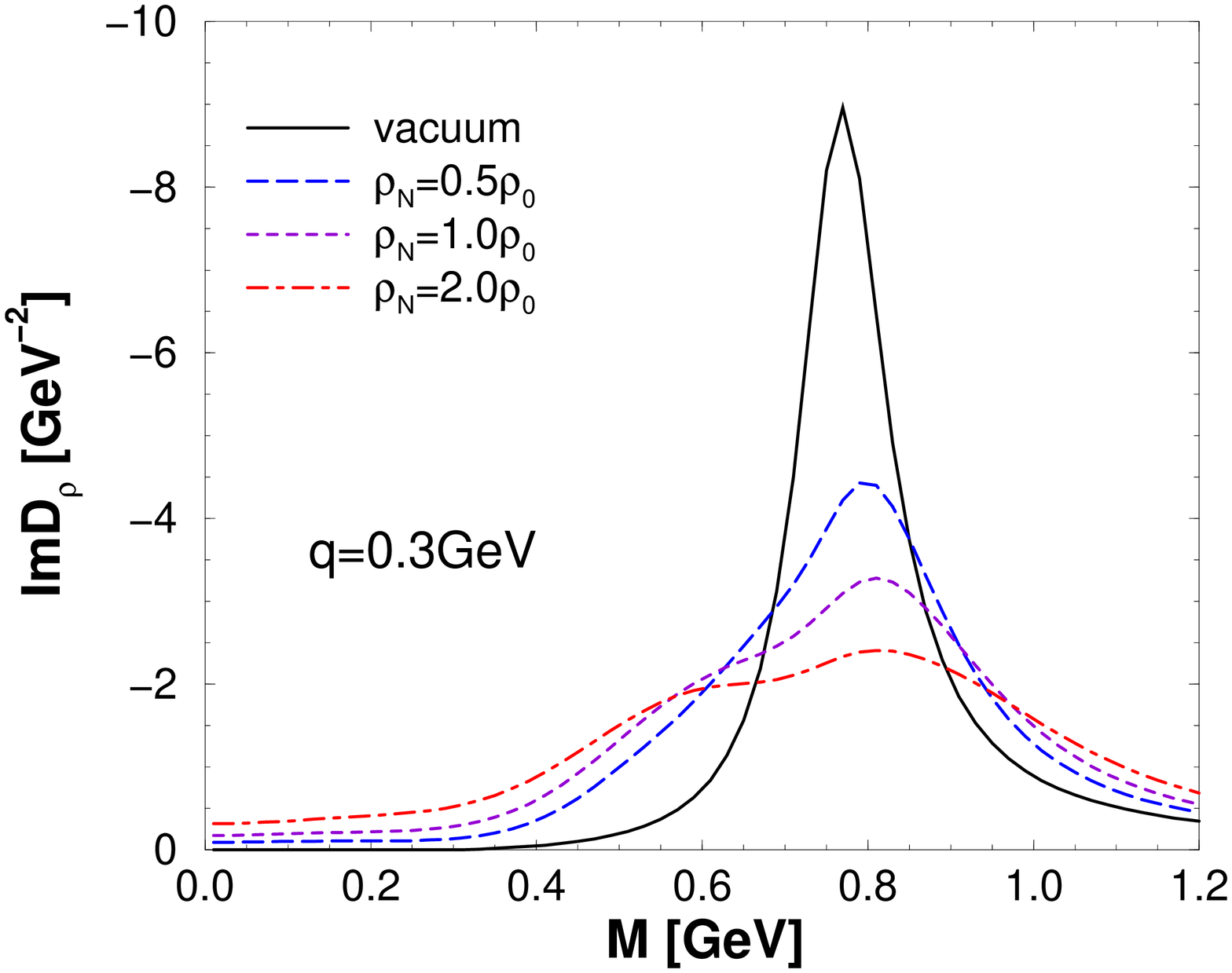,width=6.6cm,angle=-90} 
\vspace{-1.2cm}
\caption{Spin-averaged $\rho$-meson spectral function in cold nuclear 
matter~\cite{UBRW,RW99}.}
\label{fig_arho}
\end{minipage}
\end{figure}

More conservative approaches reside on many-body techniques 
to calculate selfenergy contributions to the vector-meson
propagators $D_V$ arising from interactions with surrounding matter
particles (nucleons).  
They are computed from gauge-invariant (vector-current
conserving) as well as chirally symmetric Lagrangians. 
The $\rho$ propagator, \eg,  takes the form
\beq
D_\rho^{L,T}(q_0,q;\varrho_N)=\left[M^2-(m_\rho^{(0)})^2-
\Sigma_{\rho\pi\pi}^{L,T}(q_0,q;\varrho_N)
-\Sigma_{\rho BN}^{L,T}(q_0,q;\varrho_N)\right]^{-1} \  
\eeq   	
for both transverse and longitudinal polarization states
(which in matter, where Lorentz-invariance is lost, differ for $q>0$). 
$\Sigma_{\rho\pi\pi}$ encodes the medium modifications
in the pion cloud (through $NN^{-1}$ and $\Delta N^{-1}$ bubbles, so-called 
'pisobars')~\cite{HFN92,CS92,AKLQ,KKW97,UBRW}, and 
$\Sigma_{\rho BN}$ stems from direct 'rhosobar' excitations of either 
$S$-wave ($N(1520)N^{-1}$, $\Delta(1700)N^{-1}$, $\dots$) or $P$-wave type 
($\Delta N^{-1}$, $N(1720)N^{-1}$, $\dots$)~\cite{FP97,RCW97,PPLLM,RUBW}.
The parameters of the interaction vertices (coupling constants
and form factor cutoffs) can be estimated from free decay
branching ratios of the involved resonances or more comprehensive
scattering data (\eg, $\pi N\to \rho N$~\cite{Fr98}, 
or $\gamma N$ absorption) which determine 
the low-density properties of the spectral functions.
Additional finite-density constraints can be obtained from the analysis
of photoabsorption data on nuclei. Invoking the VDM, the total
photoabsorption cross section can be readily related to the 
imaginary part of the in-medium vector-meson selfenergy in the zero
mass limit (needed for the coupling to real photons). An example
of such a calculation is displayed in Fig.~\ref{fig_phabs} where 
a reasonable fit to existing data on various nuclei has been achieved.   
\vspace{-0.5cm}
\vspace{-0.5cm}
The low-density limit (represented by the long-dashed line in 
Fig.~\ref{fig_phabs}) cannot reproduce the disappearance of
especially the $N$(1520) as seen in the data. A selfconsistent 
calculation to all orders in density~\cite{PPLLM}, however,  
generates sufficiently large in-medium widths, on the order of 
$\Gamma_{N(1520)}^{med}\simeq$~200-300~MeV (resulting in the full line). 

Fig.~\ref{fig_arho} shows the final result for the $\rho$ spectral
function~\cite{RW99} which has been subjected to the 
aforementioned constraints.  The apparent strong broadening is consistent  
with other calculations~\cite{KKW97,PPLLM}. Similar features,
albeit less pronounced, emerge within analogous treatments for 
$\omega$ and $\phi$ mesons~\cite{KKW97}.  

Let us now return to the question what these findings might imply for 
chiral restoration. In a recent work by Kim \etal~\cite{KRBR00}
an effective chiral Lagrangian including $a_1$-meson degrees of freedom
has been constructed.
Medium modifications of the latter are introduced by an '$a_1$-sobar'
through $N(1900)N^{-1}$ excitations to represent  
the chiral partner of the $N(1520)N^{-1}$ state. 
Pertinent (schematic) two-level models have been employed 
for both the $\rho$ and $a_1$ spectral densities which, in turn, have been  
inserted into the Weinberg sum rule, eq.~(\ref{wsr1}) (supplemented 
by perturbative high energy continua).
\vspace{-1.2cm}
\begin{figure}[!htb]
\bce
\epsfig{file=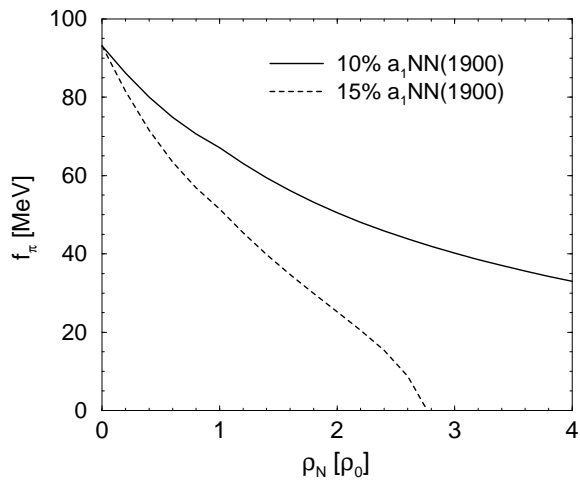,width=7cm,angle=-90}
\ece
\vspace{-0.9cm}
\caption{Pion decay constant at finite density when evaluated through
the Weinberg sum rule using schematic two-level models for
in-medium $\rho$ and $a_1$ spectral functions~\cite{KRBR00}.}
\label{fig_fpi}
\end{figure} 
\vspace{-0.5cm}
The resulting density-dependence of the pion decay constant,
displayed in Fig.~\ref{fig_fpi}, exhibits an appreciable decrease
of $\sim$~30\% at $\varrho_N=\varrho_0$, which bears some sensitivity on  
the assumed branching ratio of the $N(1900)\to Na_1$ decay (or $N(1900)Na_1$
coupling constant). However, the mechanism 
is likely to be robust: due to the low-lying $\rho$-$N(1520)N^{-1}$
and $a_1$-$N(1900)N^{-1}$ excitations, accompanied by a broadening of
the elementary resonance peaks, the $\rho$ and $a_1$
spectral densities increasingly overlap, thus reducing $f_\pi$.

\section{Electromagnetic Observables in Heavy-Ion Reactions}

In central collisions of heavy nuclei at (ultra-) relativistic 
energies (ranging from $p_{lab}$=1-200~AGeV in current experiments to
$\sqrt s$=0.2-10~ATeV at RHIC and LHC) 
hot and dense hadronic matter is created over extended 
time periods of about 20~fm/c. Local thermal equilibrium is probably 
reached within the first fm/c, after which the 'fireball' expands
and cools until the strong interactions cease ('thermal freezeout') and 
the particles stream freely to the detector. 
Electromagnetic radiation  (real and virtual photons) is continuously 
emitted as it decouples from the strongly interacting matter at the 
point of creation.  

\begin{figure}[!thb]
\vspace{-0.9cm}
\epsfig{file=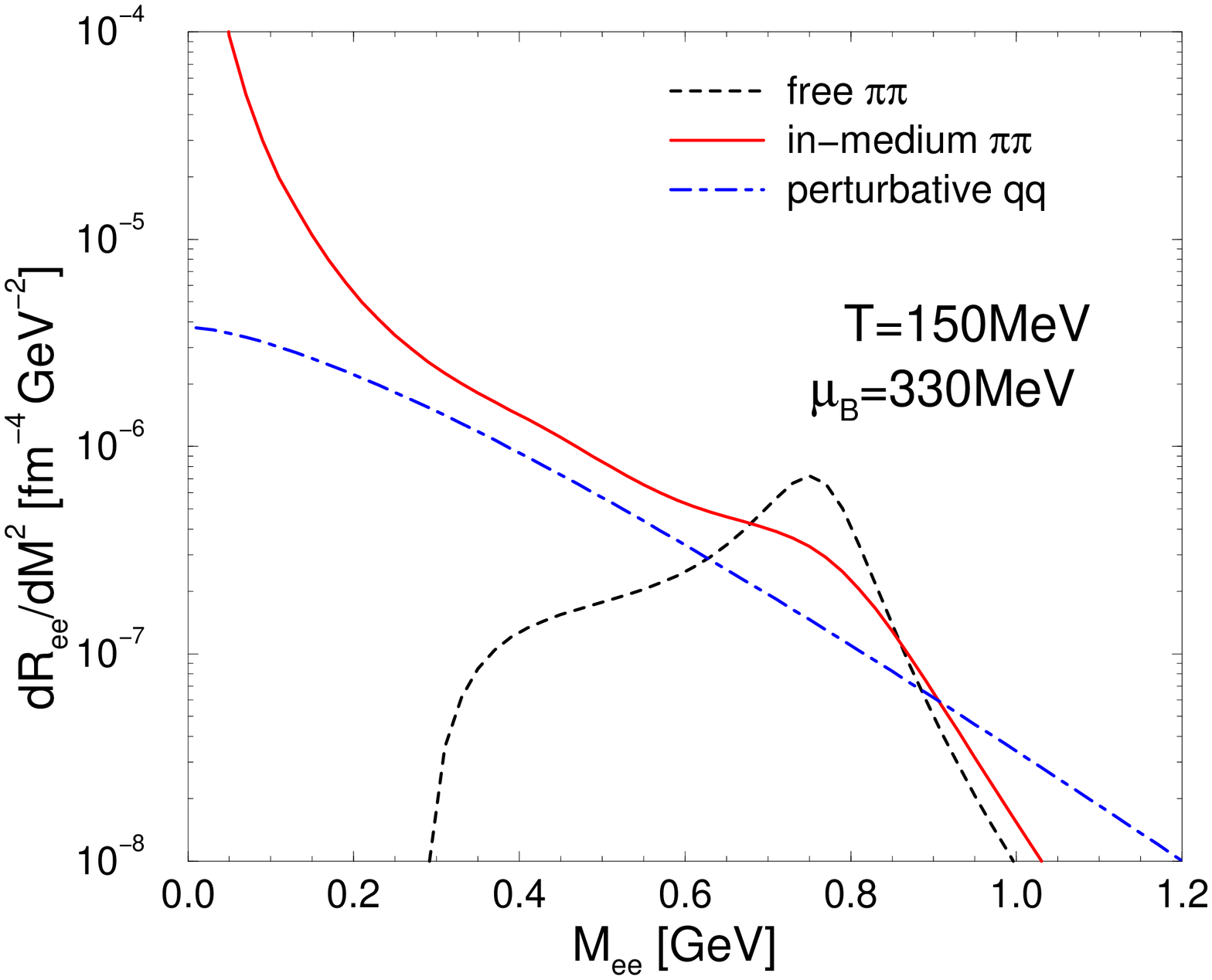,width=6.6cm,angle=-90}
\epsfig{file=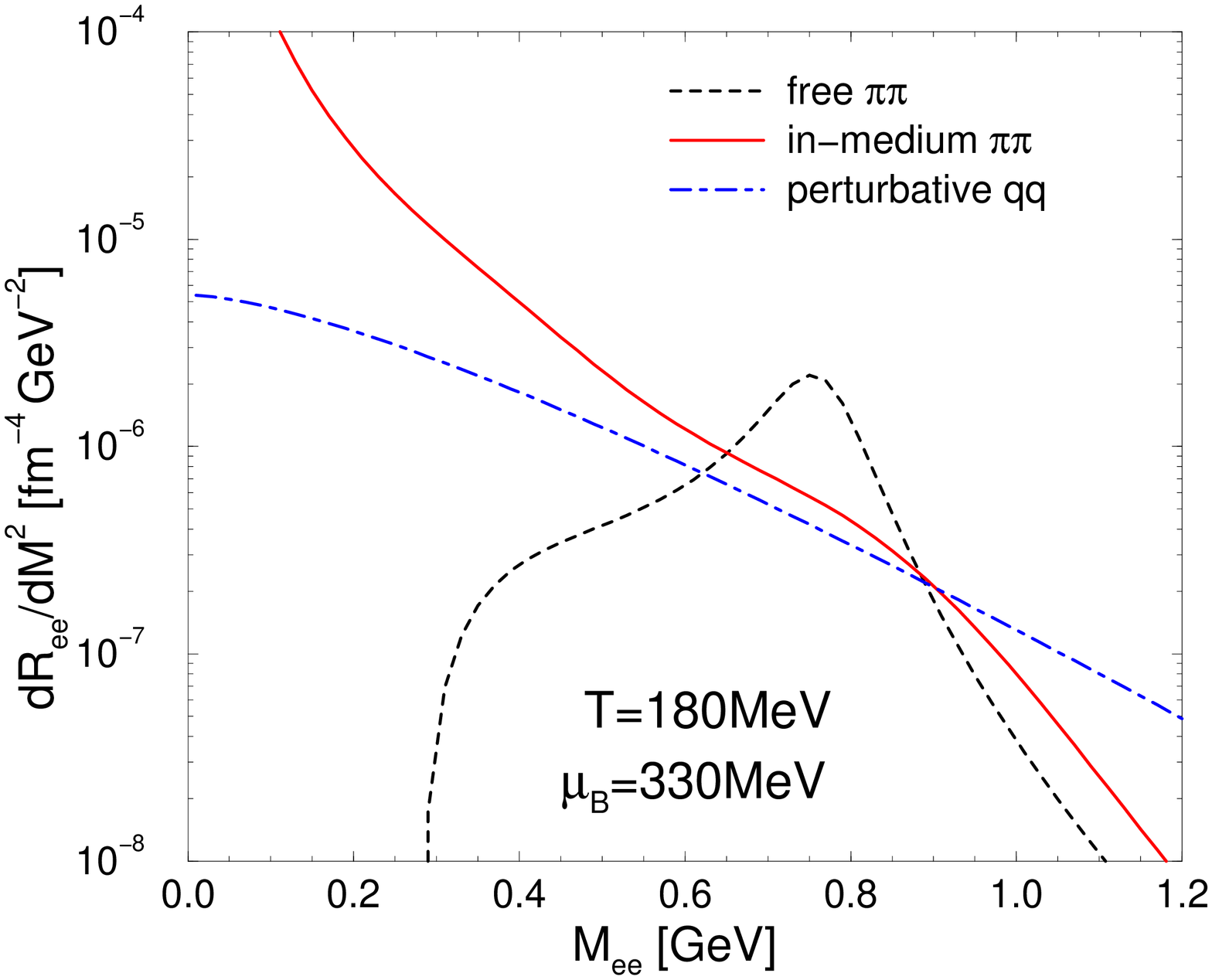,width=6.6cm,angle=-90}
\caption{$e^+e^-$ production rates in hot hadronic matter from free
$\pi^+\pi^-$ (dashed line), in-medium $\pi^+\pi^-$ (full line)~\cite{RW99}
and ${\cal O}(\alpha_S^0)$ $q\bar q$ annihilation  (dashed-dotted line).}
\label{fig_rates}
\vspace{-0.5cm}
\end{figure} 
The thermal production rate of $e^+e^-$ pairs per unit 4-volume can be 
expressed through the electromagnetic current correlation function 
(summed over all isospin states $I$=0,1), 
\beq
\frac{dN_{ee}^{th}}{d^4xd^4q}=-\frac{\alpha^2}{\pi^3 M^2} \ f^B(q_0;T)
\ \frac{1}{3} \ ({\rm Im}~\Pi_{em}^L+2 {\rm Im}~\Pi_{em}^T) 
\label{rate}
\eeq
($f^B$: Bose distribution function; 
a similar expression holds for photons with $M\to 0$). 
Fig.~\ref{fig_rates} shows that the medium effects 
in the $\rho$ propagator (including interactions with nucleons as well 
as thermal pions, kaons, etc.) induce
a substantial reshaping of the emission rate (full lines) 
as compared to free $\pi\pi$ annihilation (dashed line) 
already at rather moderate temperatures and densities 
(left panel). 
In fact, under conditions close to the expected phase boundary
(right panel) the $\rho$ resonance is completely 'melted' and the hadronic
dilepton production rate is very reminiscent to the one from a 
perturbative Quark-Gluon Plasma (dashed-dotted lines in Fig.~\ref{fig_rates})
down to rather low invariant masses
of $\sim$~0.5~GeV ($\alpha_S$ corrections to the partonic rate might
improve the agreement at still lower masses). It has been 
suggested~\cite{RW99}
to interpret this as a lowering of the in-medium 
quark-hadron duality threshold as a consequence of the approach towards
chiral restoration. 
  
\begin{figure}[!bht]
\vspace{-0.5cm}
\begin{minipage}[!t]{77mm}
\hspace{-1.0cm}
\epsfig{file=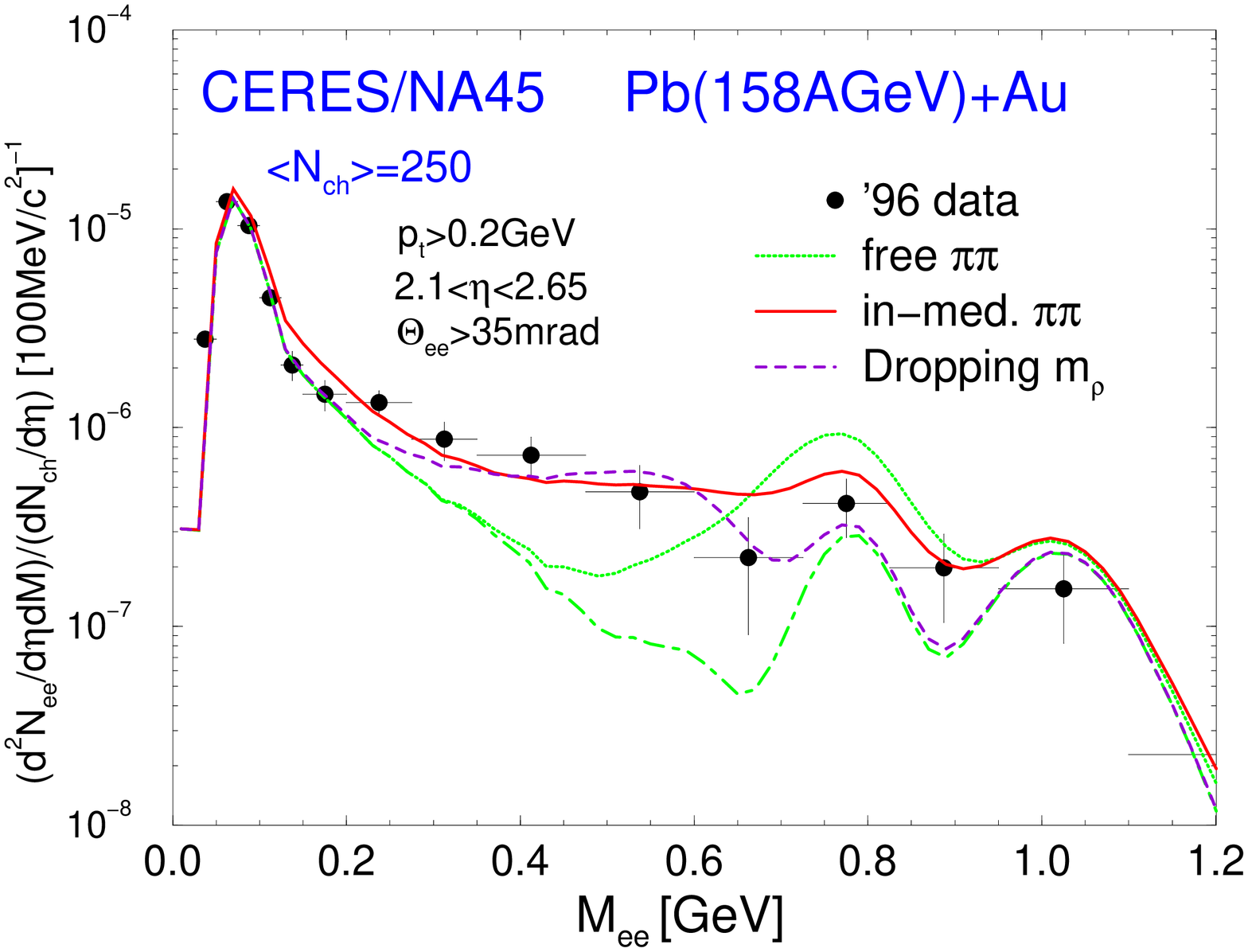,width=7cm,height=8.3cm,angle=-90}
\vspace{-1.0cm}
\caption{Low-mass dilepton spectra at the CERN-SpS~\cite{ceres}; 
Dashed-dotted line: particle decays after freezeout (hadronic 'cocktail'),
other lines: cocktail + $\pi\pi$ annihilation as indicated.} 
\label{fig_dlspec}
\end{minipage}
\vspace{-0.2cm} 
\hspace{\fill}
\vspace{-0.5cm}
\begin{minipage}[!t]{77mm}
\vspace{0.4cm}
\hspace{-0.7cm}
\epsfig{file=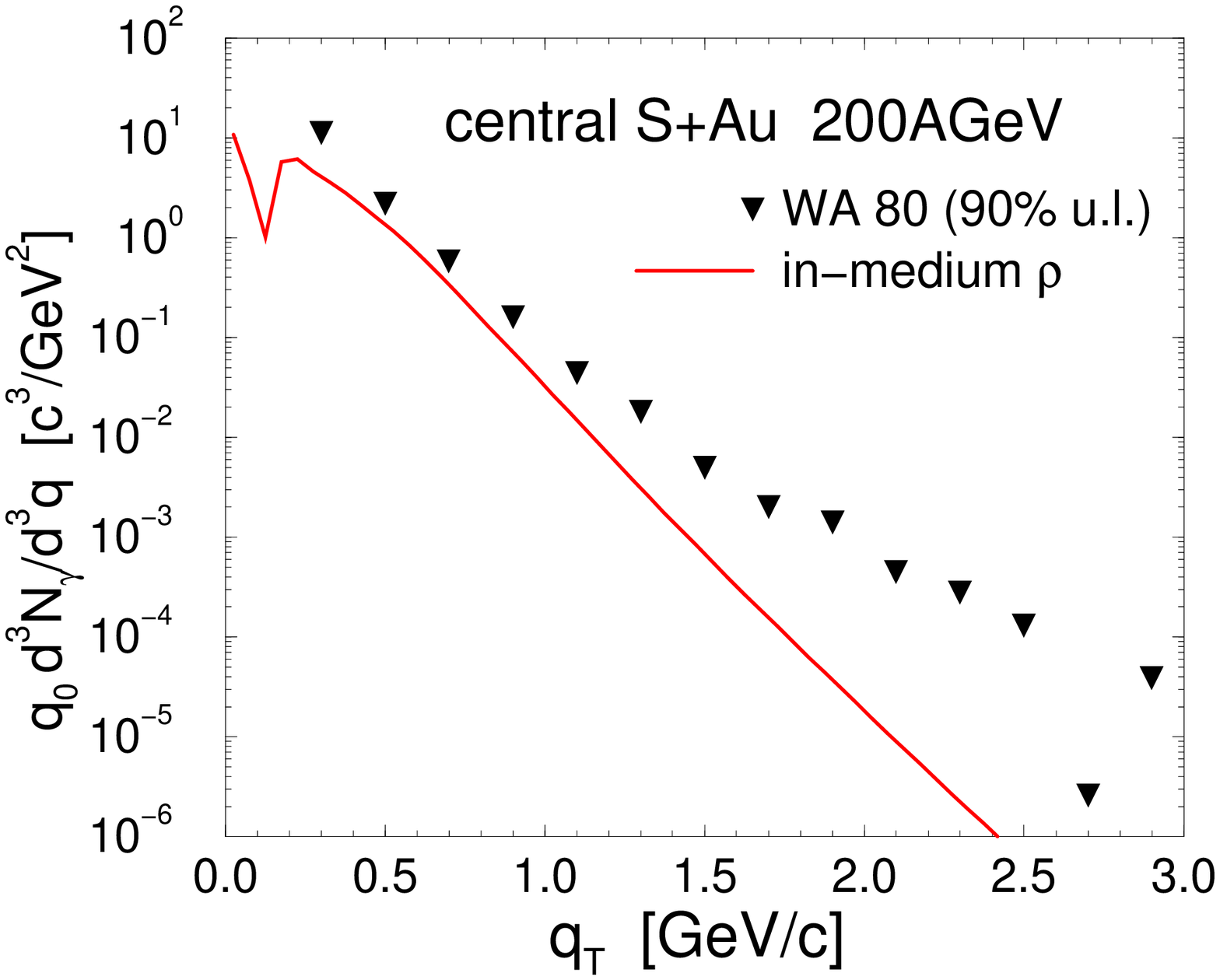,width=8cm,height=6.7cm}
\vspace{-1.2cm}
\caption{Direct photon spectra at the CERN-SpS: upper limits as extracted 
from the WA80 experiment~\cite{wa80} compared to 
in-medium thermal radiation contributions (full line)~\cite{RW00}.}
\label{fig_phspec}
\end{minipage}
\vspace{-0.2cm}
\end{figure} 
The total thermal yield in a heavy-ion reaction is obtained
by a space-time integration of eq.~(\ref{rate}) over the 
density-temperature profile
for a given collision system, modeled, \eg, within transport~\cite{CB99}  
or hydrodynamic~\cite{hydro} simulations. At CERN-SpS energies 
(160--200~AGeV) this 'thermal' component
is dominant over (or at least competitive with) final state hadron decays
(at low $M$) and hard initial processes such as Drell-Yan annihilation 
(at high $M$)
in the invariant mass range $M\simeq$~0.2-2~GeV. A consistent description
of the measured data~\cite{wa80,ceres,helios3,na50} is possible once 
hadronic many-body effects are included~\cite{RW99,RS99,GKP}, 
cf.~Figs.\ref{fig_dlspec} and~\ref{fig_phspec}.    
However, at this point also the dropping mass scenario~\cite{BR91} is 
compatible with the data~\cite{Li} (cf.~dashed curve in 
Fig.~\ref{fig_dlspec}). 

Optimistically one may conclude that strongly interacting matter
close to the hadron-QGP phase boundary has been observed at the CERN-SpS. 
Other observables such as hadro-chemistry~\cite{pbm} or 
$J/\Psi$ suppression~\cite{Satz99} also support this scenario. 
Nonetheless, further data are essential to substantiate the present 
status and resolve the open questions.

\section{Conclusions}

This talk has focused on medium modifications of vector mesons
in connection with chiral symmetry restoration in hot/dense matter. 
In accordance with a variety of empirical information 
hadronic spectral functions are characterized by the  
appearance of low-lying excitations as 
well as a broadening of the resonance structures. 
A schematic treatment of the $a_1$ meson
on similar footings shows that these features encode 
an approach towards chiral 
restoration in nuclear matter as signaled by the decrease of the pion 
decay constant when evaluating the first Weinberg sum rule.   

The application of these model calculations to electromagnetic 
observables as measured in recent heavy-ion experiments at the CERN-SpS
leads to a reasonable description of the data from 0 to 2~GeV in 
invariant mass. The structureless in-medium hadronic dilepton production 
rates resemble perturbative $q\bar q$ annihilation in the vicinity  
of the expected phase boundary indicating that chiral restoration
might be realized through a reduction of the quark-hadron duality
threshold which in vacuum is located around 1.5~GeV.  
It would also corroborate the interrelation between temperature/density
and momentum transfer in the transition from hadronic to partonic degrees
of freedom.  

In the near future further dilepton data will be taken    
by the PHENIX experiment~\cite{Phenix} at RHIC (advancing to a new energy 
frontier) as well as the precision experiment HADES~\cite{Friese} at GSI.  
Thus electromagnetic observables can be expected to continue 
the progress in our understanding of strong interaction physics.  

\vspace{1cm}

\noindent
{\bf Acknowledgments}\\
It is a pleasure to thank G.E. Brown, E.V. Shuryak and H. Sorge  
for collaboration and many fruitful discussions.


\begin{thebibliography}{9}

\bibitem{RW00}
R. Rapp and J. Wambach, 
to appear in Adv. Nucl.  Phys. (2000), and {\sf hep-ph/9909229}. 

\bibitem{Krip98}
B. Krippa, Phys. Lett. {\bf B427} (1998) 13.

\bibitem{DEI90}
M. Dey, V.L. Eletsky and B. Ioffe, Phys. Lett. {\bf B252} (1990) 620.

\bibitem{CDER98}
G. Chanfray, J. Delorme, M. Ericson and M. Rosa-Clot, {\sf nucl-th/9809007}.

\bibitem{SYZ}
J.V. Steele, H. Yamagishi and I. Zahed, Phys. Rev. {\bf D56} (1997) 5605.

\bibitem{KKW97}
F. Klingl, N. Kaiser and W. Weise, Nucl. Phys. {\bf A624} (1997) 527. 

\bibitem{LPM98}
S. Leupold, W. Peters and U. Mosel, Nucl. Phys. {\bf A628} (1998) 311. 

\bibitem{Wei67}
S. Weinberg, Phys. Rev. Lett. {\bf 18} (1967) 507. 

\bibitem{KS94}
J.I. Kapusta and E.V. Shuryak, Phys. Rev. {\bf D49} (1994) 4694. 

\bibitem{BR91}
G.E. Brown and M. Rho, Phys. Rev. Lett. {\bf 66} (1991) 2720.

\bibitem{HFN92}
M. Herrmann, B. Friman and W. N\"orenberg, Nucl. Phys. {\bf A560} (1993) 411.

\bibitem{CS92}
G. Chanfray and P. Schuck, Nucl. Phys. {\bf A555} (1993) 329. 

\bibitem{AKLQ}
M. Asakawa, C.M. Ko, P. L\'evai and X.J. Qiu, Phys. Rev. {\bf C46} (1992) 
R1159.

\bibitem{UBRW}
M. Urban, M. Buballa, R. Rapp and J. Wambach, Nucl. Phys. {\bf A641} (1998)
433. 

\bibitem{FP97}
B. Friman and H.J. Pirner, Nucl. Phys. {\bf A617} (1997) 496.

\bibitem{RCW97}
R. Rapp, G. Chanfray and J. Wambach, Nucl. Phys. {\bf A617} (1997) 472.

\bibitem{PPLLM}
W. Peters \etal, Nucl. Phys. {\bf A632} (1998) 109. 

\bibitem{RUBW}
R. Rapp, M. Urban, M. Buballa and J. Wambach, Phys. Lett. {\bf B417}
(1998) 1. 

\bibitem{Fr98}
B. Friman, in Proc. of ACTP Workshop
on 'Hadron Properties in Medium' (Seoul, Korea, 27.-31.10.97); 
and {\sf nucl-th/9801053}. 

\bibitem{RW99}
R. Rapp and J. Wambach, Eur. Phys. J. {\bf A} in press, and
{\sf hep-ph/9907502}; 
R. Rapp, Proc. of Quark Matter~'99 (Torino, Italy, 10.-15.05.99),
to appear in Nucl. Phys. {\bf A}, and {\sf hep-ph/9907342}. 

\bibitem{KRBR00}
Y. Kim, R. Rapp, G.E. Brown and M. Rho, {\sf nucl-th/9912061}. 

\bibitem{CB99}
W. Cassing and E.L. Bratkovskaya, Phys. Rep. {\bf 308} (1999) 65.

\bibitem{hydro}
J. Sollfrank \etal,  Phys. Rev. {\bf C55} (1997) 392;
C.M. Hung and E.V. Shuryak, Phys. Rev. {\bf C56} (1997) 453.

\bibitem{wa80}
R. Albrecht \etal, WA80 collaboration, Phys. Rev. Lett. {\bf 76} (1996) 3506.

\bibitem{ceres}
G. Agakichiev \etal, CERES collaboration, Phys. Lett.
{\bf B422} (1998) 405; 
B. Lenkeit, Doctoral Thesis, University of Heidelberg, 1998.

\bibitem{helios3}
A.L.S. Angelis \etal~(HELIOS-3 collaboration), Eur. Phys. J.
{\bf C5} (1998) 63.

\bibitem{na50}
E. Scomparin {\it et al.} (NA50 collaboration), J. Phys. {\bf G25} (1999) 235; 
P. Bordalo \etal~(NA50 collaboration), Proc. of Quark Matter~'99 (Torino, 
Italy, 10.-15.05.99), to appear in Nucl. Phys. {\bf A}. 

\bibitem{RS99}
R. Rapp and E.V. Shuryak, Phys. Lett. {\bf B} in press, and 
{\sf hep-ph/9909348}.

\bibitem{GKP}
K. Gallmeister, B. K\"ampfer and O.P. Pavlenko, {\sf hep-ph/9909379}. 

\bibitem{Li}
G.Q. Li, C.M. Ko, G.E. Brown and H. Sorge, Nucl. Phys. {\bf A611}
(1996) 539; G.Q. Li and C. Gale, Phys. Rev. {\bf C58} (1998) 2914.  

\bibitem{pbm}
P. Braun-Munzinger and J. Stachel, Nucl. Phys. {\bf A638} (1998) 3c.

\bibitem{Satz99}
H. Satz, Proc. of Quark Matter '99 (Torino, Italy, 10.-15.05.99), to appear 
in Nucl. Phys. {\bf A}, and {\sf hep-ph/9908339}. 

\bibitem{Friese}
J. Friese, these proceedings.

\bibitem{Phenix}
see, \eg, the PHENIX homepage at {\sf http://www.phenix.bnl.gov}.

\end{thebibliography}
\end{document}